\documentclass[times, 11pt]{article}
\usepackage{moreverb}
\usepackage{lipsum}

\newcommand{\overbar}[1]{\mkern 1.5mu\overline{\mkern-1.5mu#1\mkern-1.5mu}\mkern 1.5mu}

\usepackage[colorlinks,bookmarksopen,breaklinks,bookmarksnumbered,hyperindex,citecolor=red,urlcolor=red]{hyperref}
\usepackage{setspace}
\newcommand\BibTeX{{\rmfamily B\kern-.05em \textsc{i\kern-.025em b}\kern-.08em
T\kern-.1667em\lower.7ex\hbox{E}\kern-.125emX}}

\usepackage{amsmath}

\RequirePackage[OT1]{fontenc}
\usepackage{array}

\usepackage{multirow}
\usepackage{multicol}
\usepackage{pdflscape}
\usepackage{afterpage}
\usepackage{caption}
\usepackage[margin=1.05in]{geometry}
\usepackage{booktabs}
\usepackage{rotating}
\usepackage{parskip}
\usepackage{color,soul}
\usepackage{graphicx}
\graphicspath{%
    {converted_graphics/}
    {/}
    {C:/Users/user/Documents/}
}
\begin{document}

\title{Estimating the average treatment effects of nutritional label use using subclassification with regression adjustment}






\raggedright
\section{Introduction}       

\indent Disclosure of ingredients and inclusion of a standardized label has been required on all US food and beverage since 1994 as a result of the National Labeling Education Act (NLEA, \cite{labeling1990education}). The United States Food and Drug Administration \cite{FDAcite} initially estimated, among other benefits, roughly 725,000 avoided cases of cancer and chronic heart disease over a 20-year period and a health care savings between \$4.4 and \$26.5 billion through expected dietary changes resulting from the NLEA.

Twenty years later, however, the effect of the NLEA on health outcomes remains largely unknown, as literature exploring the effect of label use has yielded mixed conclusions. While Variyam and Cawley \cite{variyam2006nutrition} and Loureiro $et$ $al.$ \cite{loureiro2012effects} found a significant reduction in body mass index (BMI) among women label users, Drichoutis $et$ $al.$ \cite{drichoutis2005nutrition} found evidence that increased label use actually caused higher BMI's.  However, both Variyam and Cawley and Loureiro $et$ $al.$ dichotomized label use, initially measured on a 5-pt scale, into `sometimes' frequency or above, making inference at a specific label use level impossible. Further, in dichotomizing an ordered exposure, both studies were more likely to suffer from bias due to confounded assignment mechanism \cite{royston2006dichotomizing}. 

Estimating the effects of an exposure on an ordinal scale is useful for many public health interventions. For example, extensive clinical trials have contrasted the duration, length, and intensity levels of physical activity (including \cite{chambliss2005exercise, puetz2008randomized}, to name a few). Such research has aided in proposing recommendations for physical activity, including those touted by the U.S. Surgeon General \cite{united1996physical}. Obviously, these guidelines cannot be enforced, however they were written in order to motivate people to live healthier lifestyles, and to identify the average effects that are expected due to different activity levels. 

Similar guidelines on how often one should read nutritional labels have not been issued, despite label use being a priority for several United States organizations. The U.S. Food and Drug Administration \cite{FDA}, the American Heart Association \cite{AHA}, and the American Diabetes Association \cite{ADA}, for example, all include label use directions on their websites. The Mayo Clinic \cite{mayo} goes as far as urging patients to `practice' label use when food shopping. However, none of these organizations supply any specific guidelines of how often individuals should be reading nutritional labels.  

Observational data that use a simple comparison of health outcomes across those at different label use levels has limitations, because subjects in these label use groups differ with regard to personal, socio-economic and demographic characteristics. For example, readers of nutrition labels are, on average, more active and health-conscious \cite{neuhouser1999use, satia2005food}. With two treatment groups, a common statistical tool used to adjust for differences in the covariates' distribution in estimation of the treatment effect is the propensity score, defined by Rosenbaum and Rubin \cite{rosenbaum1983central} as the probability of receiving treatment conditional on a set of observed covariates. Most propensity score methods and applications deal with binary treatments, while exposure to label use is often measured using an ordinal scale. In the 2005-06 National Health and Nutrition Examination Survey (NHANES) data, label use is measured on a five-point scale,  $never$, $rarely$, $sometimes$, $most$ $of$ $the$ $time$ ($often$), or $always$. Drichoutis $et$ $al.$ \cite{drichoutis2005nutrition} employed binary propensity score methods across the ten possible pairs of label levels in their analysis of the NHANES, which yielded pairwise causal effect estimates that were not transitive. Specifically, estimates suggest that while $sometimes$ label use level yields lower BMI than $rare$ level and that $rare$ level causes lower BMI than $never$ level, $sometimes$ level frequency actually results in a significantly higher BMI ($p$-value \textless 0.05) than \emph{never} level.
 
We extend the data set used by Drichoutis $et$ $al.$ \cite{drichoutis2005nutrition} and reanalyze it with a generalized propensity score method that will result in transitive estimates of the causal effects of increased label use on BMI between all pairs of label use levels. In doing so, this manuscript provides three important extensions to approaches which have been previously designed for ordinal exposures \cite{joffe1999invited, imai2004causal}. Following the separation of the design and analysis paradigm in observational studies proposed by  \cite{rubin2008objective}, we propose and implement novel graphical methods as well as introduce new metrics for assessing and depicting covariates' similarity between individuals at different exposure levels. Second, we couple the subclassification based strategies of \cite{imai2004causal} and \cite{zanutto2005using} with regression adjustments to estimate causal effects and to obtain more precise and accurate point estimates. Lastly, we use simulations to demonstrate the benefits of combining subclassification with regression adjustment, relative to either method alone and to other previously proposed methods for ordinal exposures. Although previous statistical literature has touched on some of the analysis phase methods, the combination of the design, simulation, and analysis phases presented here provide other investigators a complete case study for estimating causal effects from observational studies with ordinal treatments. Our method is implemented on the 2005-06 NHANES, and causal effect estimates suggest that reduction in BMI only occurs when reading labels $often$ or $always$. 

The outline of this paper is as follows.  Section 2 introduces our notation, and Section 3 details our use of subclassification with regression adjustment to estimate the set of causal effects across levels of an ordinal exposure. Section 4 implements the proposed method on the NHANES data, Section 5 summarizes our results, and Section 6 details a simulation study. Section 7 concludes.   

\section{Causal Inference \& the Rubin Causal Model}
		
\subsection{Notation for binary treatment}
Neyman \cite{splawa1990application} first described treatment effects in the context of potential outcomes for a randomized experiment.  This concept was expanded to observational studies in what was eventually termed the `Rubin Causal Model' (RCM) \cite{holland1986statistics, rubin1974estimating}. 

Let $Y_i$, $\boldsymbol{X}_i$, and $T_i$ be the observed outcome, set of $p$ covariate values, and binary treatment indicator, respectively, for each subject \emph{i = 1,... n}, $n \textless N$, where $n$ is the sample size and $N$ is the population size which is possibly infinite, with treatment $T_i \in \mathcal{T}$, $\mathcal{T} = \left\{0,1 \right\}$. 

A commonly made assumption in the RCM is the stable unit treatment value assumption (SUTVA) \cite{rubin1980randomization}. SUTVA specifies both that the set of potential outcomes for a subject depends only on the treatment that subject was assigned to, and not on the treatment assignment of others, and that within each treatment condition, there are not multiple versions of the treatment. Assuming SUTVA, the potential outcome for unit $i$ can be written as $Y_i(T_i=t)=Y_i(t)$, which represents subject $i$'s outcome if he or she would have received treatment $t$. 

One common estimand of interest is the population average treatment effect ($PATE$), which is often approximated by using the sample average treatment effect ($SATE$). \begin{eqnarray}
	PATE &=&  E[Y(1)-Y(0)] \label{pat}  \\
	SATE &=& \frac{1}{n}\sum_{i=1}^n (Y_i(1)-Y_i(0)) \label{sat} \end{eqnarray} In practice, however, each individual receives either the treatment or the control at the same point in time, but not both, and only $Y_i(1)$ or $Y_i(0)$ is observed for each unit, also known as the fundamental problem of causal inference \cite{holland1986statistics}. As a result, the RCM commonly relies on the assumption S1 to estimate \ref{pat} and \ref{sat}. 

\begin{description}
\item[S1:] Strongly ignorable treatment assignment: (i) $Pr(\left\{Y(0), Y(1) \right\}|T , \boldsymbol{X})$ = $Pr(\left\{Y(0), Y(1) \right\}|\boldsymbol{X})$ and (ii) $0 \textless Pr(T=t|\boldsymbol{X}) \text{ for } t \in \left\{0,1\right\}$ \cite{rosenbaum1983central}. Under strongly ignorable treatment assignment, the set of potential outcomes and treatment assignment are conditionally independent given $\boldsymbol{X}$. Implicit in this assumption is that differences in outcomes between those with the same $\boldsymbol{X}$ are unbiased estimates of the treatment's causal effect to units with that $\boldsymbol{X}$.  
\end{description}

To estimate causal effects from observational data, matching subjects with the same $\boldsymbol{X}$ who received different treatments is an effective way of reducing bias, but as the dimension of $\boldsymbol{X}$ increases, this is nearly impossible \cite{stuart2010matching}. Propensity scores enable inference under the RCM even in a high dimensional setting. Let $e(\boldsymbol{X}) = Pr(T=1|\boldsymbol{X})$ be the propensity score. If treatment assignment is strongly ignorable given $\boldsymbol{X}$, then it is also strongly ignorable given $e(\boldsymbol{X})$, $Pr(\left\{Y(0), Y(1) \right\}|T,e(\boldsymbol{X})) = Pr(\left\{Y(0), Y(1) \right\}|e(\boldsymbol{X}))$. Thus, the comparison of units with equal $e(\boldsymbol{X})$'s is unbiased for estimating unit level effects, and averaging over the distribution of $e(\boldsymbol{X})$ in the population results in an unbiased estimate of the $PATE$ \cite{rosenbaum1983central}.  

\subsection{Expansions for more than two exposure levels}

Assuming SUTVA, for $Z$ exposures or exposure levels, with $\mathcal{T} = \left \{1...Z\right \}$, let $\mathcal{Y}_i = \left \{Y_i(1), Y_i(2), ..., Y_i(Z)\right \}$, where $\mathcal{Y}_i$ is the set of potential outcomes for unit $i$. With an ordinal exposure, possible estimands of interest are the population average treatment effects between exposure levels $t$ and $s$, $PATE_{t,s}$, for all pairs $\left \{t,s\right \}$, where $t,s \in \mathcal{T}$, which are commonly approximated by the sample average treatment effects, $SATE_{t,s}$. \begin{eqnarray} PATE_{t,s} &=&  E[Y(t)-Y(s)]   \\
SATE_{t,s} &=& \frac{1}{n}\sum_{i=1}^n (Y_i(t)-Y_i(s))\end{eqnarray} As with binary treatment, we cannot observe each $SATE_{t,s}$ because each unit only receives one treatment, and therefore $SATE_{t,s}$ is a random quantity due to the assignment mechanism being random. Assuming that the sample is randomly chosen from the population, then $SATE_{t,s}$ is an approximation for the $PATE_{t,s}$. Because most applications are usually trying to estimate effects generalizing to the population, from this point forward we will define $PATE_{t,s}$ as our estimand of interest, and assume that the observed data was sampled at random from the population.

To estimate the $PATE$ across exposure pairs, S1 is expanded such that a strongly ignorable treatment assignment mechanism (also called strong unconfoundedness) for multiple exposures states that (i) $Pr[\mathcal{Y}|T=t, \boldsymbol{X}] = Pr[\mathcal{Y}|\boldsymbol{X}]$ and (ii) $0\textless Pr[T=t|\boldsymbol{X}] \; \forall \; t \in \mathcal{T}$. As in the binary treatment setting, SUTVA and a strongly ignorable treatment assignment mechanism enable us to estimate $E[Y(t)]$, for all $t$, by conditioning on the observed covariates. 

The propensity score has been expanded to multiple exposures through the generalized propensity score (GPS), $r(t,\boldsymbol{X}) = Pr(T=t|\boldsymbol{X=x})$ \cite{imai2004causal, imbens1999role, lechner2001identification}. While propensity scores for binary treatment enable us to condition on a scalar in order to estimate treatment effects, the GPS with a discrete exposure may consist of multiple dimensions, thus requiring to condition on an entire vector of treatment assignment probabilities, $\boldsymbol{r(X)} = (r(1,\boldsymbol{X}),...,r(Z,\boldsymbol{X}))$. As a result, two individuals with the same $r(t,\boldsymbol{X})$ for one specific treatment level may not be equivalent with regards to their entire $\boldsymbol{r(X)}$. Thus, differences in outcomes between subjects with different exposure levels and similar $r(t,\boldsymbol{X})$, but differing $\boldsymbol{r(X)}$, are not generally unbiased causal effect estimates \cite{imbens1999role}.

Joffe and Rosenbaum \cite{joffe1999invited} and Imai and van Dyk \cite{imai2004causal} noted that modeling an ordinal exposure using an ordered logit model, also referred to as the proportional odds model \cite{mccullagh1980regression}, can provide a shortcut to conditioning on a multidimensional $\boldsymbol{r(X)}$. The ordered logit model is appropriate for exposures measured in doses (e.g., low, medium, high). For example, with $Z$ total treatments (exposure levels), assuming		\begin{eqnarray}  
			 \text{log}\bigg(\frac{P(T_i \textless t)}{P(T_i \geq t)}\bigg) = \theta_t - \boldsymbol{\beta^TX_i}, t = 1,...,Z-1 \label{MCC} \end{eqnarray} and defining the balancing score, $b(\boldsymbol{X})$, as a function of the covariates such that $Pr(T=t|b(\boldsymbol{X}))$ = $Pr(T=t|b(\boldsymbol{X}),\boldsymbol{X})$, the proportional odds model provides a scalar $b(\boldsymbol{X}$).  Specifically, for $\boldsymbol{\beta^T} = (\beta_1, .. \beta_p)^T$, $\boldsymbol{\beta^TX}$ is a balancing score, such that  \begin{eqnarray} Pr(T=t|\boldsymbol{\beta^TX}) = Pr(T=t|\boldsymbol{X}, \boldsymbol{\beta^TX)} \text{ for $t = 1, .... Z$} \label{balancescore} \end{eqnarray} 
The combination of Result (\ref{balancescore}) with the assumption of a strongly ignorable treatment assignment mechanism allows us to establish that $\mathcal{Y}$ and the treatment assignment are conditionally independent given $\boldsymbol{\beta^TX}$ (for a proof, see Imai and van Dyk \cite{imai2004causal}), 
\begin{eqnarray} Pr[\mathcal{Y}|T=t,\boldsymbol{\beta^TX}] = Pr[\mathcal{Y}|\boldsymbol{\beta^TX}] \label{StrongIgnore} \end{eqnarray}

\noindent Under the expanded versions of SUTVA and S1, differences in observed $Y$'s between subjects with different exposure levels but equal $\boldsymbol{\beta^TX}$ are unbiased estimates of causal effects at that $\boldsymbol{\beta^TX}$. To estimate the $PATE_{t,s}$ for all treatment pairs $\left \{t,s\right \}$, we want to average $E[Y(t)-Y(s)|\boldsymbol{\beta^TX}]$ over the distribution of $\boldsymbol{\beta^TX}$. Formally, we would estimate $E[Y(t)-Y(s)]$ using the following: \begin{eqnarray} E[Y(t)-Y(s)] &=& E[E(Y(t)-Y(s)|\boldsymbol{X})] \nonumber \\
			&=& E[E(Y(t)-Y(s)|\boldsymbol{X},\boldsymbol{\beta^TX})] \nonumber \\
			&=& E[E(Y(t)-Y(s)|\boldsymbol{\beta^TX)}] \nonumber \qquad \qquad \qquad   \qquad \qquad \text{ (by Result (\ref{balancescore}))}\\	\label{EY}&=& \int (E[Y(t)|T=t,\boldsymbol{\beta^TX}]-E[Y(s)|T=s,\boldsymbol{\beta^TX}])Pr[\boldsymbol{\beta^TX}] 
	d(\boldsymbol{\beta^TX}) \label{int} \end{eqnarray} \noindent Direct computation of (\ref{int}), however, is difficult because it requires integrating over the probability distribution of $\boldsymbol{\beta^TX}$. 

One approach for approximating the $PATE_{t,s}$ is to partition subjects with similar values of $\boldsymbol{\beta^TX}$ into subclasses, estimating the effect within each subclass, and combining these effects using a weighted average. A second alternative could be the use of radius matching \cite{caliendo2008some} to pair subjects with roughly equivalent $\boldsymbol{\beta^TX}$'s, and average across pairs. However, individual matching techniques are not as well suited for multiple treatments \cite{imbens1999role}. A third approach, which is discussed in Section \ref{Alternative}, uses inverse probability weighting. 

\section{Subclass weighted causal effects for an ordinal exposure}

\subsection{Design phase}

Estimation of causal effects using observational data is composed of two phases: the design phase and the analysis phase \cite{rubin2001using}. The design phase is done without the outcome in sight, and with the intent of obtaining the same treatment effects which would have been obtained in a completely randomized design \cite{rubin2008objective}. As suggested by Joffe and Rosenbaum \cite{joffe1999invited} and implemented by Lu $et$ $al.$ \cite{lu2001matching}, we first use Equation (\ref{MCC}) to fit $Pr(T|\boldsymbol{X})$ and generate an estimated $\boldsymbol{\hat{\beta}X_i}$ for each individual, where $\boldsymbol{\hat{\beta}}$ is the maximum likelihood estimate of $\boldsymbol{\beta}$. The goal in the design phase is to group subjects which are similar with respect to the observed covariates \cite{rubin2008objective}. Thus, we are not concerned with assessing the fit of treatment assignment (e.g., testing the proportional odds assumption), but whether balance on all covariates is obtained across treatment groups.

\subsubsection{Covariate choice}

The choice of which covariates to include in the generalized propensity score model should be made with the intent of satisfying the assumption of strong ignorability. Primarily, previous scientific research should be used to instruct choice of $\boldsymbol{X}$ \cite{rubin2001using}, with all measured pre-treatment variables associated with both the treatment assignment and the outcome included \cite{rubin1996matching}. In addition, when in doubt, Stuart \cite{stuart2010matching} recommends a `liberal' inclusion variables associated with either the treatment assignment or the outcome, because exclusion of variables which are associated with the treatment assignment mechanism can increase bias. 

While it cannot be verified that the chosen $\boldsymbol{X}$ satisfies the assumption of strong ignorability, Stuart \cite{stuart2010matching} argues that strong ignorability is often more valid than it appears because controlling for observed covariates also controls for correlated but unobserved ones. As part of the covariate selection, we propose to examine if any covariates which were not included in the treatment assignment model are also balanced across subclasses. Exact implementation will be described in Section~\ref{Res}. 

\subsubsection{Common support}

As with propensity score analysis for binary treatment, it is important to eliminate subjects outside the range of common support \cite{dehejia1998causal}. With binary treatment, a common support is often considered to be the range of propensity scores of those receiving both treatments. For an ordinal exposure, an extension is to use a common support region of the linear predictor, which eliminates subjects with $\boldsymbol{\hat{\beta}^TX}$ beyond the range of $\boldsymbol{\hat{\beta}^TX}$ values among those on other treatments. It is recommended that the propensity score model be re-fit after subjects are dropped to ensure that the estimated propensity scores are not disproportionately impacted by those outside the common support \cite{ImbensRubinBook}. Dropping units also changes the estimand of interest to include only units with a large enough probability of receiving any of the treatments. This is a different estimand than the $PATE_{t,s}$, which cannot be estimated without making unassailable assumptions. Thus, it is good practice to describe the population which the estimand is generalizable to, using the observed covariates.  

The remaining subjects that are not discarded are partitioned into $K$ subclasses, where each subclass contains subjects with similar $\boldsymbol{\hat{\beta}^TX}$. This partitioning is aimed at generating similar covariates' distributions for all treatment levels in each subclass. The choice of $K$ is flexible, and it has been suggested to examine the covariate balance for multiple values of $K$ \cite{rubin2001using}. Higher $K$ will yield better within-subclass homogeneity of the covariates, resulting in smaller within-subclass bias. Too large of a $K$ will result in low numbers of subjects within each subclass, which could restrict our ability to estimate causal effects when there are no units at a specific treatment level to compare to. \hl{For simplicity, we partition units into subclasses such that an equal number of units are within each subclass. Cochran and Rubin} \cite{cochran1973controlling} \hl{found little improvement when comparing the bias reduction of optimal subclassification to equally spaced subclassification with a single covariate and a binary treatment, and Rosenbaum and Rubin} \cite{rosenbaum1984reducing} \hl{provided similar recommendations when estimating the treatment effect with multiple covariates and binary treatment. Our recommendation is to use equally spaced subclasses with ordinal treatments and multiple covariates, but this is an area of further research.}

Let $n_k$ be the number of subjects in subclass $k$, $k = 1,...,K$. With binary treatment and $p$ covariates in the propensity score model, it has been recommended to keep (i) at least three subjects at each combination of the subclass and treatment and (ii) $n_k \textgreater p+2$ \cite{ImbensRubinBook}. Our related recommendation is to generate the largest $K$ possible with both (i) at least $3+Z$ subjects at each exposure level in each subclass and (ii) $n_k \textgreater p+Z$. 

\subsubsection{Balance checks}

To ensure that subclassification reduced the covariates' bias across the different treatment groups, it is important to check the within-subclass distributions of each covariate before looking at within-subclass outcomes \cite{rubin2001using, rubin2008objective}. This process examines how closely each subclass mimics a randomized experiment in which the distributions of covariates at each exposure level are similar in expectation. 

The following two-step procedure was used to examine the covariate distributions within each subclass. First, tabular and graphical approaches assess the distributions of both $\boldsymbol{\hat{\beta}^TX}$ and the continuous covariates in $\boldsymbol{X}$ by exposure level within each subclass \cite{austin2009balance}. These checks include side-by-side boxplots of the balancing scores and continuous variables at each exposure level in each subclass. 

Second, the dependencies between exposure level and covariate within each subclass, for all covariates, will be compared to both the dependencies in the original data and the hypothetical distribution of the statistics which would have occurred in a randomized experiment. Here, we use Kendall's $\tau_b$, abbreviated as $\tau$ from this point forward, which is a rank correlation coefficient, where positive $\tau$ values indicate that higher ranks of one covariate are positively associated with higher ranks of the exposure. Under the null distribution that the covariate and exposure are independent, $\tau=0$, and sample $\tau$ statistics are approximately distributed as standard Normal, making $\tau$ useful for examining non-linear correlations. We plot histograms of sample $\tau$ test statistics for each covariate at each subclass to check for normality, as well as to identify the proportion of $\tau$ statistics which remain significant after subclassification, relative to nominal level $\alpha$.
 
Examining all of the $\tau$ values for each covariate in each subclass may be extensive with a large number of covariates. One way to summarize the benefits of subclassification is to average the within-subclass $\tau$ estimates for each variable over the number of subclasses, and compare these results to the values found in the original data. Formally, let $\tau_{pk}$ be the estimated $\tau$ between exposure level and covariate $p$ in subclass $k$, and let $w_k= \frac{n_k}{n}$ be the proportion of subjects in subclass $k$. We define $\overbar{\tau}_{p}$, the weighted subclass-averaged $\tau$, as \begin{eqnarray} \overbar{\tau}_{p} = \sum_{k=1}^K \tau_{pk} w_k \nonumber \end{eqnarray} Contrasting the $\overbar{\tau}_{p}$ values with the $\tau$ statistics from the original data can indicate if covariate imbalances still exist. 

Section~\ref{BA} details these checks through real data analysis. If these checks display covariate imbalances which deviate from a randomized experiment, one option would be to re-fit the ordered logistic model, possibly including interaction terms. Noticeable variations in the distributions of  $\boldsymbol{\hat{\beta}^TX}$ or significant $\tau$ dependencies within each subclass, for example, would suggest that the covariates are not properly balanced. If balance on $\boldsymbol{X}$ cannot be obtained, causal effects should not be calculated.

\subsection{Analysis phase}

Under strong ignorability, if the empirical distribution of the covariates is equal in expectation between those at different exposure levels within each subclass, estimated mean outcomes for each treatment level can be computed as weighted averages of the within-subclass sample means, with weights equal to the relative subclass size. Let $\bar{y}_{kt}$ and $\bar{y}_{ks}$ be the observed sample means in subclass $k$ among those receiving treatments $t$ and $s$, respectively. To test for a global difference in subclass-weighted mean outcomes between the exposure levels, Zanutto $et$ $al.$ \cite{zanutto2005using} use a randomized block ANOVA model of outcome on subclass and exposure, treating subclass as the blocking variable. If the global difference in means hypothesis is rejected, pairwise  $PATE_{t,s}$'s can be estimated using subclass weighted mean differences, as in Equation~(\ref{three}).\begin{eqnarray}  \widehat{PATE}_{(t,s)} &=& \sum_{k=1}^K ( \bar{y}_{kt} w_k - \bar{y}_{ks} w_k) \label{three} \label{Zan}\end{eqnarray}

Without regression adjustment, however, subclass weighted means may not eliminate the entire bias caused from differences in the covariates' distribution, jeopardizing the accuracy of treatment effects estimated using Equation (\ref{three}). The intuition behind this is that while differences in outcomes are unbiased estimates of causal effects at exact values of the linear predictor, differences in covariates by exposure level could still exist when different linear predictors are pooled together. Several authors (\cite{rubin1979using, lunceford2004stratification, abadie2006large}, to name a few) have noted that combining regression adjustment with matching for a binary treatment reduces bias relative to either method alone. An additional benefit of regression adjustment is that even in the case that the theoretical covariate balance of a completely randomized design is achieved within each subclass, regression adjustment can improve the precision of the causal estimates \cite{ImbensRubinBook}. 

We start the analysis by testing for a global effect of exposure using a randomized block ANCOVA model of outcome on subclass, exposure, and $\boldsymbol{X}$, treating subclass as the blocking variable. If the null hypothesis of no difference in means by exposure is rejected, we calculate pairwise causal effects.

Let $Y_{ik}$ be the observed outcome of subject $i$ in subclass $k$, and let $Y_{ik}(t)$ be the potential outcome of that subject at exposure level $t$. Next, letting $\boldsymbol{X_{ik}}$ be the observed covariates of subject $i$ in subclass $k$, and $I(T_i=t)$ be an indicator function for individual $i$ receiving treatment $t$, we use the following steps to estimate $PATE_{(t,s)}$ for all pairs $\left \{t,s \right \}$.  

\begin{description}
\item[Step 1] Assuming $Y_{ik}(t)|\boldsymbol{X_{ik}} \sim N(E(Y_{ik}|X_{ik},T),\sigma^2)$, model $Y_{ik}|\left\{\boldsymbol{X_{ik}},T\right\}$ within each subclass using the following regression model  \end{description} \begin{eqnarray}  E(Y_{ik}|X_{ik},T) &=& \sum_{t=1}^Z \alpha_{kt}I_t(T_i=t) + \boldsymbol{{\gamma_k}X_{ik}} \label{RM}\\
	&=& \alpha_{k1}I(T_i=1) + ... + \alpha_{kZ}I(T_i=Z) + \boldsymbol{{\gamma_k}X_{ik}}  \nonumber \end{eqnarray}
\begin{description}
\item[Step 2] Estimate $PATE_{k(t,s)}$, the $PATE_{(t,s)}$ within subclass $k$, using $\hat{\alpha}_{kt}$ and $\hat{\alpha}_{ks}$, the maximum likelihood estimates of $\alpha_{kt}$ and $\alpha_{ks}$, respectively, from Model (\ref{RM}): \end{description} \begin{eqnarray} \widehat{PATE}_{k(t,s)} &=& \hat{\alpha}_{kt} - \hat{\alpha}_{ks} \label{eqn} \end{eqnarray}
\begin{description}
\item[Step 3] Estimate the variance of $\widehat{PATE}_{k(t,s)}$, $Var(\widehat{PATE}_{k(t,s)})$, within each subclass, from regression model (\ref{RM})\end{description} 
	Let $\hat{\boldsymbol{\alpha}}_k' = (\hat{\alpha}_{k1},  ... ,\hat{\alpha}_{kZ})$ with $Var(\hat{\boldsymbol{{\alpha}}}_k) =\boldsymbol{\widehat{\Sigma}_k}$. Based on (\ref{RM}), $\hat{\boldsymbol{\alpha}}_k \sim \mathcal{N}(\boldsymbol{\alpha_k},\boldsymbol{\Sigma_k)}$, and letting $\boldsymbol{c} = (\boldsymbol{0}, I(T=t), \boldsymbol{0},-I(T=s),\boldsymbol{0})$, where $I(T=t)$ and $I(T=s)$ are indicators for treatments $t$ and $s$, respectively, with $\boldsymbol{0} = (0, ... , 0)$, we have
\begin{eqnarray} Var(\widehat{PATE}_{k(t,s)}) = Var(\hat{\alpha}_{kt} - \hat{\alpha}_{ks})  = Var(\boldsymbol{c\hat{\alpha}}_k) = \boldsymbol{c\widehat{\Sigma}_k c'} \label{eqnvar} \end{eqnarray}
\begin{description}
\item[Step 4] Using $w_k= \frac{n_k}{n}$, estimate $PATE_{t,s}$ by averaging over $K$:\end{description}\begin{eqnarray} 
\widehat{PATE}_{(t,s)} &=& \sum_{k=1}^{K} w_k (\widehat{PATE}_{k(t,s)}) \label{eqn2} \\ 
 \widehat{SE}(\widehat{PATE}_{(t,s)}) &=& \sqrt{\sum_{k=1}^{K} w_k^2 (\widehat{Var}(\widehat{PATE}_{k(t,s)}))} \label{eqn6} \end{eqnarray}

\noindent Using our framework, $\hat{\alpha}_{kt}- \hat{\alpha}_{ks}$, the estimated average treatment effect between level $t$ and $s$ in subclass $k$, is an unbiased estimate for $PATE_{k(t,s)}$ (For proof, see Appendix \ref{proof}). It is important to note that because $n_k$ and the linear predictors are both based on the GPS model estimated from the data, responses within and between subclasses are dependent \cite{du1998valid}. As a result, the above aggregation of subclass weighted standard errors can underestimate the true sampling variances, although regression adjustment usually helps in this regard \cite{du1998valid, benjamin2003does}.

\subsection{Alternative Approaches}
\label{Alternative}
In addition to subclassification based methods, other inference procedures exist for estimating causal effects from an ordinal exposure. Lu $et$ $al.$ \cite{lu2001matching} used non-bipartite matching to pair subjects at lower exposure levels with ones at higher levels. However, the causal effect estimand generated using non-bipartite matching is not clearly defined, and a significant effect using this method would not specify an optimal exposure level. 

The approach used by Drichoutis $et$ $al.$ \cite{drichoutis2005nutrition}, initially described by Lechner \cite{lechner2002program}, is also common for estimating treatment effects from multiple exposures. Letting $n_t$ be the number of subjects receiving treatment $t$, this method implements a set of binary comparisons (SBC) attempting to estimate the population average treatment effect on the treated, $PATT_{t|(t,s)}$ = $E[Y(t)-Y(s)|T=t]$, for all exposure pairs $\left \{t,s \right \}$, using propensity score matching for binary treatment on the population of subjects receiving either $t$ or $s$. Because SBC yields causal effects conditional on a subject receiving one of two treatments, the resulting set of causal effects are usually not transitive. Specifically, the population receiving $t$ which $PATT_{t|(t,s)}$ generalizes to likely differs from the population receiving $s$ which $PATT_{s|(s,r)}$ generalizes to, and, as a result, it would be erroneous to use $PATT_{t|(t,s)}$ and $PATT_{s|(s,r)}$ to contrast treatments $r$ and $t$. 

Another approach for approximating the $PATE$ between each exposure pair uses the inverse of the estimated probabilities from a statistical model of treatment assignment (e.g. multinomial logistic, proportional odds) as weights \cite{imbens1999role, mccaffrey2013tutorial}. Feng $et$ $al.$ \cite{feng2012generalized} used this procedure to estimate $PATE_{t,s}$ by weighting subjects by the reciprocal of their GPS. \begin{eqnarray}\widehat{PATE_{t,s}} &=&\widehat{E[Y(t)]} -\widehat{E[Y(s)]} \text{ where} \label{Feng} \\
 \widehat{E[Y(t)]} &=& \Bigg( \sum_{i=1}^n \frac{I(T_i=t) Y_i}{r_{(t,\boldsymbol{X_i})}}\Bigg) \Bigg(\sum_{i=1}^n \frac{I(T_i=t)}{r_{(t,\boldsymbol{X_i})}}\Bigg)^{-1} \text{ and} \nonumber  \\
 \widehat{E[Y(s)]} &=& \Bigg( \sum_{i=1}^n \frac{I(T_i=s) Y_i}{r_{(s,\boldsymbol{X_i})}}\Bigg) \Bigg(\sum_{i=1}^n \frac{I(T_i=s)}{r_{(s,\boldsymbol{X_i})}}\Bigg)^{-1} \nonumber \end{eqnarray}

One issue with this approach is that extreme weights can result in erratic causal estimates \cite{little1988missing, kang2007demystifying}, an issue which becomes more likely as the number of treatments increases and treatment assignment probabilities decrease. While trimming has been shown to decrease the influence of extreme weights on causal estimates \cite{huber2013performance}, trimming the extreme weights estimated from a GPS model can yield covariate bias' in unknown directions \cite{kilpatrick2012exploring}. 

Nonetheless, our subclassification estimators can be viewed as weighted estimators, with weights coarsened by averaging them through subclasses. For binary treatment, this smoothing of the weights results in estimates which, compared to weighted methods, are more precise and less likely to be influenced by a misspecification of the propensity score model \cite{ImbensRubinBook, stuart2008best}. 

\section{Nutritional label use and BMI}

\subsection{Data description}
\label{Sec4}
The NHANES is a nationally representative research program of 15 United States counties that measures demographic, health, nutritional, and behavioral variables, including nutritional label use and BMI. The 2005-06 NHANES version measured label use via a questionnaire and BMI through a physical examination. Subjects were presented with an example of a food label and asked the question
\begin{center} \emph{`How often do you use the Nutrition Facts panel when deciding to buy a food product?  Would you say always, most of the time, sometimes, rarely, or never?'}\footnote{ See 
\href{http://www.cdc.gov/nchs/data/nhanes/nhanes\_05\_06/sp\_dbq\_d.pdf}{\texttt{http://www.cdc.gov/nchs/data/nhanes/nhanes\_05\_06/sp\_dbq\_d.pdf}} for more information }\end{center}
 \noindent In a separate physical examination, trained medical personnel measured the height and weight of these subjects. 

Thirty pre-treatment covariates that are possibly associated with label use exposure and BMI, including demographic, lifestyle, nutritional awareness, and health status information, were chosen after careful examination of the NHANES survey and a vast literature review (\cite{neuhouser1999use,lewis2009food}, to name a few). All of the variables recommended by Drichoutis $et$ $al.$ \cite{drichoutis2005nutrition} were  included. We added squared terms for \emph{Metabolic equivalence} and \emph{Meals away from home}  to account for the skewed nature of the original variables \cite{rubin2001using}. The covariate $Weight$ $thoughts$, which measures an individual's categorized opinion of their weight (underweight, about the right weight, or overweight), was also included. Lastly, we included the variable \emph{Prior BMI}, which is calculated using a self reported estimate of a subject's weight from a year prior to the survey and the subject's current measured height.

The data set included a total of 4,644 subjects with recorded label use and a measured BMI. As in Drichoutis $et$ $al.$ \cite{drichoutis2005nutrition}, we excluded the 298 subjects with missing covariates values. Including \emph{Prior BMI} as a covariate eliminated an additional 74 subjects, yielding a sample size of 4,272. Because dealing with missing covariates is not the focus of this paper, we made the naive assumption that data for these subjects were missing completely at random \cite{rubin1978bayesian}. Other options include introducing \emph{missing} indicators for categorical covariates \cite{d1998tutorial}, using weighting methods based on the probability for missingness (as in  \cite{wooldridge2007inverse}), or using multiple imputations to create complete data sets, where causal effect estimates are calculated across each of the data sets and combined using Rubin's rules for multiple imputation \cite{rubin1996multiple}. Because these techniques have not yet been used with GPS methods under multiple exposure levels, it is an important area for further research. Selected demographic variables of subjects dropped using this criteria and those remaining in the study population are shown in Appendix \ref{Compare}.

\vspace{5 mm}
\begin{center}***************************Table \ref{Kendall} here***************************\end{center}
\vspace{5 mm}

Table~\ref{Kendall} lists our covariates, their $\tau$ statistics with label use, and a $p$-value testing the null hypothesis of no dependency between label use and each covariate.\footnote{There are 33 rows in Table 1, as we separated the variable for race into four categories. For a more complete description of these covariates, see Appendix \ref{NHANES})} Using these covariates, the ordered logistic model was used to estimate the probability of label use (the treatment).

\subsection{Balance assessment} \label{BA}

Subjects were partitioned into $K$ equal size subclasses, with subclass boundaries defined by \hl{equally spaced} quantiles of $\boldsymbol{\hat{\beta}^T X}$. There were 33 covariates in the propensity score model. To meet the restrictions of (i) at least $3+Z$ subjects at each label use level within each subclass and (ii) $n_k \textgreater p + Z$, up to $K$ = 15 subclasses were examined. Balance checks are presented for $K$ = 5, 10, and 15.

\subsubsection{Distributions of $\boldsymbol{\hat{\beta}^T X}$ and balance checks for continuous covariates} 

Boxplots of $\boldsymbol{\hat{\beta}^T X}$ by label use within each subclass show that while the linear predictors are distributed similarly among those at different label use levels for $K=10$ and $K=15$, those with higher label use levels have higher $\boldsymbol{\hat{\beta}^T X}$ within each subclass for $K=5$. For example, in subclass 4 with $K=5$, the boxplots indicate a pattern of increasing $\boldsymbol{\hat{\beta}^T X}$ by label use level (Figure 1). However, when these subjects are further split on $\boldsymbol{\hat{\beta}^T X}$, as in subclasses 7 and 8 with $K=10$, the linear predictor appears more evenly distributed across label use levels (Figure 1).

\vspace{5 mm}
\begin{center} ***************************Figure 1 here***************************\\ 
\vspace{5 mm}
\begin{em} \footnotesize
Boxplots of $\boldsymbol{\hat{\beta}^T X}$ (the linear predictor) by label use in subclass 4 ($K$=5) and subclasses 7 and 8 ($K$=10) \end{em}\normalsize \end{center}
\vspace{5 mm}

Overlap and similarities in the distributions of continuous covariates by label use were also compared via side by side boxplots, both overall and within each subclass. Extreme continuous covariates' values may have large influence on the causal estimates, particularly if the overlap of continuous variables is not roughly equal across label use levels. One option is to perform the analysis on a common support of continuous variables, by eliminating subjects whose covariates are beyond the range of those at other label use levels. For example, sample cutoff lines used with this inclusion criteria for the variable \emph{Prior BMI} are shown in Figure 2, which eliminated, along other subjects, a subject with a \emph{Prior BMI} of 87.5. This elimination was done before the propensity score model was estimated, and would be done prior to any elimination of extreme linear predictors. Another option was to exclude subjects with extreme continuous variables within each subclass, but in the NHANES data set, this would eliminate more than 30\% of the participants, and thus this strategy was not attempted. Elimination changes the population for whom the results can be generalized to, but it reduces the need for extrapolation and making assumptions which cannot be defended. 

\vspace{5 mm}
\begin{center}***************************Figure 2 here***************************\\ 
\vspace{5 mm}
\begin{em} \footnotesize
Boxplots of Prior BMI by label use, with cutoffs for `extreme' values \end{em}\normalsize \end{center}
\vspace{5 mm}

\subsubsection{Within-subclass associations between $\boldsymbol{X}$ and $T$ using Kendall's $\tau$} As an example of balance assessment using $\tau$, let \emph{Drug user} be a binary variable for whether or not a subject indicated using hashish, marijuana, cocaine, heroin, or methamphetamine in the past 12 months. One significant sample $\tau$ statistic occurred with \emph{Drug user} in subclass 2, for $K$ = 10 (Table~\ref{druguser}). In this example, $\tau$ = 0.09, suggesting an increase in label use is associated with an increase in the likelihood of using drugs, as the $z$-statistic for this association is 2.00. 

\vspace{5 mm}
\begin{center} ***************************Table \ref{druguser} here*************************** \end{center}
\vspace{5 mm}

With several hundred such tests, however, we expected to find these associations by chance, as well. Figure~4 in Appendix~\ref{figsec} depicts the distributions of the test statistics plotted against a normal curve, and Table~\ref{taustab} shows the proportion of significant tests observed after subclassification at level $\alpha$, $\alpha \in \left \{0.01, 0.05\right \}$. In Figure 4, we look for normality in the histograms, and in Table~\ref{taustab}, because the distribution of p-values is uniform under the null, we check that the proportion of significant tests is near $\alpha$. Results are presented across three choices of $K$ for the following three mechanisms of subject elimination, E1-E3:

\begin{description}
\item[E1:] No subject elimination, n = 4272
\item[E2:] Eliminate subjects with extreme linear predictors, n = 4142
\item[E3:] Eliminate subjects with extreme continuous $X$ or extreme linear predictors, n = 4076
\end{description}

\vspace{5 mm}
\begin{center} ***************************Table \ref{taustab} here***************************\end{center}
\vspace{5 mm}

\noindent These checks show that while there were significant within-subclass covariate imbalances beyond that which would have occurred in a randomized design when $K$ = 5, the proportion of significant tests of dependency dropped for $K$ = 10 and $K$ = 15. The variables $Age$, \emph{Drug user}, \emph{Healthy diet},  \emph{Heard of food guide pyramid}, \emph{Pregnant}, \emph{Prior BMI}, \emph{Weight thoughts}, and \emph{Doct. advice 3: eat less fat for disease risk} displayed the strongest ($p$-value \textless 0.05) tests of within-subclass dependency for $K$ = 10 and 15. 

Lastly, we compare $\tau$ statistics before any subclassification with subclass-weighted $\overbar{\tau}_{p}$ statistics, for $K$ = 5 and 15, under elimination mechanism E3 (Figure 3). This figure is an extension of the `Love' plot proposed for binary treatment, which is popular for showing post-matching decrease in each covariates' bias \cite{ahmed2006heart}. Twenty six of the 33 $|\tau|$ statistics using the original data are greater than 0.02, and 19 of these correlations are greater than 0.10. For $K= 15$, no $|\overbar{\tau}|$ is of magnitude greater than 0.016, and 29 of the 33 $|\overbar{\tau}|$ are less than 0.01. Dependencies appear to still exist within subclasses for $K=5$, where 10 $|\overbar{\tau}|$ are greater than 0.02. For $K=10$ (not shown), the largest  $|\overbar{\tau}|$ is 0.019 (\emph{Metabolic Equivalence}). 

\vspace{5 mm}
\begin{center} ***************************Figure 3 here***************************\\ 
\vspace{5 mm}
\begin{em} \footnotesize
Kendall's $\tau$ between covariates and label use, before and after stratification (using $K = \left \{5,15\right\}$) \end{em}\normalsize \end{center}
\vspace{5 mm}

These results suggest that subclassifying with $K$= 10 and 15 eliminated most of the differences in observed covariate distributions across label use categories which were found in the original data. Because our checks deem covariates to be plausibly balanced for these $K$'s only, we do not estimate within-subclass causal effects for $K=5$.  

\subsection{Subclass weighted causal effect estimates of label use on BMI with regression adjustment}

Let $BMI_{ik}(t)$ be the potential outcome BMI of subject $i$ in subclass $k$ at label use $t$, for $i = 1, ..., n$, $k=1, ..., K$, $K \in \left \{10,15 \right \}$, and $t \in \left \{ 1 = never, 2 =rare, 3 = some, 4 = \emph{ most of the time (often)}, 5 = always\right \}.$ With $Z$ = 5 and $Y_{ik}(t) = BMI_{ik}(t)$, Equations \ref{RM}-\ref{eqn6} were used to estimate the $PATE_{(t,s)}$ and their variances for all pairs $\left \{t,s\right \}$. 

Estimates for three forms of subject elimination (E1-E3) and two regression model adjustments (A1-A2) are shown in Table~\ref{ATE}. The regression adjustment models were used to adjust for lingering bias that was not eliminated using subclassification. Model A1 included the set of covariates with questionable balance as judged by within subclass $\tau$ statistics, as described in Section 4.2, and model A2 included all covariates in Table ~\ref{Kendall}.

\begin{description}
\item[A1:] $\boldsymbol{X}$ =  $Age$, \emph{Drug user}, \emph{Healthy diet},  \emph{Heard of food guide pyramid},\emph{Pregnant}, \emph{Prior BMI}, \emph{Weight thoughts}, and \emph{Doct. advice 3: eat less fat for disease risk } (See Appendix \ref{NHANES} for variable definitions)
\item[A2:] $\boldsymbol{X}$ = All covariates in Table~\ref{Kendall} 
\end{description}

\noindent 

Two other sets of causal effects are presented in Table~\ref{ATE}. First, estimates calculated using SBC, as detailed in Section \ref{Alternative} and calculated by Drichoutis $et$ $al.$ \cite{drichoutis2005nutrition} with this same data set, are displayed.\footnote{\cite{drichoutis2005nutrition} used several matching algorithms in their analysis. The estimates shown in Table~\ref{ATE} reflect those using one-to-one nearest-neighbor matching.} Second, we calculated IPTW estimates of the $PATE$'s, as in Feng $et$ $al.$ \cite{feng2012generalized} and Equation \ref{Feng}.\footnote{As in \cite{feng2012generalized}, we used bootstrap sampling to estimate the variance of the IPTW causal effects.}  

\vspace{5 mm}
\begin{center} ***************************Table \ref{ATE} here***************************\end{center}
\vspace{5 mm}

\section{Results} \label{Res}

Using a randomized block ANCOVA model with $K=10$ and $K=15$ subclasses as blocks, at the 0.05 nominal level, we rejected the global null hypothesis of no differences between the mean BMI's at each label use \hl{($p$ \textless 0.01 for both $K$, using each combination of unit discarding rule (E1-E3) and regression adjustment method (A1,A2))}. Examining the estimated $PATE$'s between the 10 pairs of label levels suggest that $often$ or $always$ label use may yield lower BMI than $rare$ or $sometimes$ usage. However, the majority of comparisons are not significant at the 0.05 level; the one comparison that was significant across most models examined suggests that an $often$ usage yields a lower BMI than a $rare$ one. Effect estimates are similar for different unit discarding rules (E1-E3), choice of $K$, and regression adjustment method (A1,A2). IPTW estimates are mostly inconclusive, save for limited evidence that $often$ levels cause lower BMI than $rare$ and $sometimes$ levels. 

The marginal increase in BMI with low levels of label use, relative to no label use, is a bit of a surprise; one possibility is that subjects who read labels at a minimum level falsely believe that they are acting sufficiently healthy, and respond with behaviors or eating habits which increase BMI.  Another possibility is that the strong ignobility assumption is violated, which implies that subjects reading at the $rare$ levels are unique in a dimension not captured by the observed covariates. However, this violation is less plausible when a large number of covariates are being balanced. 

The causal estimates provided are only unbiased under the assumptions specified in Section 2. SUTVA seems reasonable for the NHANES. However, we caution that merging label use categories into two levels (as in \cite{variyam2006nutrition, loureiro2012effects}) may violate the multiple version of treatment assignment assumption. The NHANES data also included other covariates that were not included in the GPS model because we felt that other variables served as sufficient proxies. As a sensitivity analysis, we examined six of these covariates: cocaine use, marijuana use, marital status, an indicator for excessive alcohol consumption, blood pressure problems, and desires for weight control (listed in Appendix \ref{Extra}, along with their pre-subclassification Kendall's $\tau$ with label use). Using our split of subjects into 15 subclasses, we tested for within-subclass dependency between label use and these covariates using Kendall's $\tau$. Of the 90 tests, 1 (1.1\%) and 4 (4.4\%) were significant at $\alpha = 0.01$ and $\alpha = 0.05$, respectively, roughly what would have occurred in a randomized design. Thus, it appears that we were able to balance observed covariates even when they were not explicitly included in the GPS model.

Our decision to eliminate subjects with extreme linear predictors or continuous variables (E2, E3) results in estimands that are different than $PATE$'s, and the estimates provided in Table~\ref{ATE} each generalize to different populations. However, under both E2 and E3, fewer than 5\% of subjects were eliminated. Two variables, education level and familiarity with the food guide pyramid, offered the strongest insight into why subjects were not retained. Of the 130 subjects eliminated under E2 and the 196 subjects dropped under E3, 61 had the lowest education level and had no knowledge of the food guide pyramid. An additional 46 eliminated subjects had the highest education level and were familiar with the food guide pyramid. These types of subjects were less likely to be observed at all label use levels and would require extrapolation. 

Compared to other methods for ordinal exposures applied to this data set, subclassification with regression adjustment provides important advantages. In the IPTW analysis, 309, 313, and 307 subjects were given a weight greater than 10 under E1, E2, and E3, respectively, yielding causal effects with larger variances in comparison to our proposed method. The maximum weights under the three elimination mechanisms were 129 (E1), 108 (E2), and 57 (E3). Subclassification based estimates are also transitive and generalizable to the entire study population that is not discarded, whereas estimates using a SBC generalize to separate subsets of the population and are not transitive. Here, transitivity refers to the additive effects of causal estimates across different exposure levels.  For example, using our method, but not that of a SBC, the additive effects of $often$ to $some$ and $some$ to $rare$ label use frequency is equivalent to comparing $often$ to $rare$ usage.

\section{Simulation}

In real data, true causal effects are not known because each subject receives only one treatment or exposure dose at a specific time point. If complete sets of potential outcomes were known for all subjects, however, it would be straightforward to compare competing methods to see which most accurately and precisely estimates the true $PATE$. Thus, we created two full data sets that include the full set of potential outcomes which could have occurred if we had observed the subjects at all label use levels. The two sets of full data, Set 1 and Set 2, used the 2005-06 NHANES with label use as exposure and BMI as outcome. Letting $BMI_i(t)$ be the potential outcome $BMI$ under treatment $t$ for subject $i$, we imputed two fixed sets of potential outcomes as follows:

\begin{description} 
\item[SET 1] $PATE_{(t,s)} = 0$ for all $\left \{t,s\right \}$. Here, $BMI_i(t) = BMI_i$  for all $t \in \mathcal{T}$, where $BMI_i$ is the observed $BMI$ for unit $i$ in the data set.\\	

\item[SET 2] $PATE_{(t,s)} \ne 0$ for all pairs $\left \{t,s\right \}$. Imputation of these potential outcomes were obtained using the following algorithm.

\begin{enumerate} \item The principal components of $X$, the matrix of covariates listed in Table~\ref{Kendall}, were calculated.\footnote{Here, we excluded the squared terms for \emph{Metabolic equivalence}and \emph{Meals away from home}, as the inclusion of these variables led to erratic principal components. For more information on the principal components procedure, see \cite{jolliffe2005principal}.}
\item All subjects were projected to the eigenvector ($\boldsymbol{V_1}$) that corresponded to the largest eigenvalues of $\boldsymbol{X}$, $PC1_i = \boldsymbol{V_1^T X_i}$.
\item $BMI_i(T_i)$, the potential outcome at subject $i$'s observed treatment assignment, was set as the observed outcome, $BMI_i$. 
\item For $t \neq T_i$, the potential outcomes were imputed using the observed BMI outcomes of the subjects receiving other treatment levels whose $PC1$'s were closest to that of subject $i$. Specifically, $BMI_i(t) = BMI_j(t) = BMI_j, \forall \text{ }t = T_j = T_{j'} \ne T_i$, where $|PC1_i - PC1_j| \leq |PC1_i - PC1_{j'}| \text{ } \forall \text{ }j'$.   \end{enumerate}  

\end{description}
\noindent For Set 2, the resulting population average causal effects for the different usage level comparisons were: -0.14 ($rare$ vs. $never$), -0.18 ($some$ vs. $never$), -1.20 ($often$ vs. $never$), and 0.32 ($always$ vs. $never$).

At each simulation step, we applied the following algorithm:

\begin{enumerate}
\item Randomly select 15 of the covariates listed in table 1 without replacement, and let $\boldsymbol{X_{sim}}$ be the matrix with these covariates
\item Estimate $\hat{\boldsymbol{\gamma}}_{t}$, the maximum likelihood estimate of $\boldsymbol{\gamma}$ from the multinomial logisitic regression model $log(\frac{P(T=t)}{P(T=z)}) = \boldsymbol{\gamma_{t}}\boldsymbol{X_{sim}}$ based on the observed $T$ and $\boldsymbol{X_{sim}}$.
\item Let $\hat{r}_{sim,i}(t,\boldsymbol{X_{sim}})$ be the estimated probability that unit $i$ received treatment $t$, based on the model in the previous step with $\boldsymbol{\gamma}$= $\hat{\boldsymbol{\gamma}}$, and sample $T_{sim,i}$ based on $\hat{r}_{sim,i} = (\hat{r}(1,\boldsymbol{X_{sim,i}}),\dots,\hat{r}(5,\boldsymbol{X_{sim,i}})).$
\item Set the observed outcome $BMI_{sim,i}=BMI(T_{sim,i})$.
\end{enumerate}

It is important to note that both the treatment assignment mechanism and the outcome model are different than the $GPS$ model and the linear regression model, respectively. We used $BMI_{sim}$, $T_{sim}$, and $\boldsymbol{X_{sim}}$ to compare seven methods of estimating the $PATE$ across pairs of label use dosages. The seven methods included four variations of subclassification and three commonly used comparison approaches. Subclassification techniques were generated by combining two factors, the number of subclasses used ($K$ = 5, 15) and whether or not regression adjustment for all covariates in $\boldsymbol{X}$ was used within each subclass (yes, no). The three commonly used estimation methods included the naive differences in the sample means of $BMI_{sim}$ between those at different treatment levels, $\widehat{PATE}_{t,s}$ = $\frac{1}{n_t} (\sum_{i=1}^n BMI_{(sim,i)}*I(T_{sim}=t))$ - $\frac{1}{n_s}(\sum_{i=1}^n BMI_{(sim,i)}*I(T_{sim}=s))$. The second method used standard regression adjustment of $BMI_{sim}$ on $T_{sim}$ and $\boldsymbol{X}$, with the causal effects estimated using the coefficients on $T_{sim}$. The last method relied on IPTW with normalized weights (Equation (\ref{Feng})). In this calculation, subjects receiving level $t$ were weighted by $1/(Pr(T_{sim}=t))$, where $Pr(T_{sim}=t)$ was calculated using the proportional odds model.

At each simulation $m$, $m = 1, ..., 2000$, we estimated $\widehat{PATE}_{m(t,s)}$ and its standard error, $SE(\widehat{PATE}_{m(t,s)})$, for each of the seven estimating procedures and dose comparisons. This yielded simulated bias ($bias_m$) and coverage indicators ($coverage_m, allcoverage_{m})$ for each procedure at each $m$: \begin{eqnarray} bias_{m(t,s)} &=& \widehat{PATE}_{m(t,s)} - PATE_{(t,s)} \nonumber \\
 coverage_{m(t,s)} &=& \left\{ 1 \text{ if } PATE_{t,s} \in \widehat{PATE}_{m(t,s)} \pm 1.96*SE(\widehat{PATE}_{m(t,s)}), 0 \text{ otherwise} \right \} \nonumber \\
allcoverage_{m} &=& \left\{ 1 \text{ if } coverage_{m(t,s)} = 1 \text{ for all pairs } \left \{t,s \right \}, t \ne s, 0 \text{ otherwise} \right \} \nonumber \end{eqnarray}
The mean bias, $\overline{bias}_{t,s} = \frac{1}{2000} \sum_{m=1}^{2000} bias_{m(t,s)}$, was calculated for each of the ten pairs of dose comparisons, as well as the standard deviation of bias. We present results for the four dose comparisons with $never$ label use, as results for other mean bias calculations are similar. Two summary statistics for coverage rates, $Average$ and $Complete$ coverage, are also shown for each method, where \begin{eqnarray}
Average &=& \frac{1}{20000} \sum_{m=1}^{2000} \sum_{t,s}^{10} coverage_{m(t,s)} \nonumber \\
Complete  &=& \frac{1}{2000} \sum_{m=1}^{2000} allcoverage_{m(t,s)} \nonumber \end{eqnarray}

\noindent Because we did not adjust for multiple interval estimations, $Complete$ coverage is expected to be lower than both $Average$ coverage and the nominal level. 

\vspace{5 mm}
\begin{center} ***************************Table \ref{ATESim} here***************************\end{center}
\vspace{5 mm}

Results of the simulations are depicted in Table \ref{ATESim}. Regression alone and subclassification with regression adjustment yielded the lowest $\overline{bias}$ for Set 1, $PATE_{t,s} = 0$. All of the subclassification approaches showed lower $\overline{bias}$ and higher coverage rates for Set 2, $PATE_{t,s} \ne 0$, compared to the other methods. Among the subclassification methods implemented, a higher number of subclasses and the inclusion of regression adjustment tended to yield higher coverage rates and lower $\overline{bias}$. IPTW estimates showed higher bias and lower coverage, possibly due to the misspecified treatment assignment model or the sensitivity of this procedure to large weights. With a binary treatment assignment, misspecified treatment assignment models and extreme weights can yield causal effects with larger bias and higher MSE \cite{kang2007demystifying, stuart2008best, waernbaum2012model}.

The results of our simulations suggest that when the estimated treatment assignment mechanism, in this case the proportional odds model, does not reflect the true assignment mechanism, a method involving subclassification with regression adjustment can outperform competing estimators of $PATE$ for ordinal exposures. Further, combining subclassification with regression adjustment yields lower bias and higher coverage rates when compared to either method alone.

\section{Discussion}

The analysis presented here adds to that of Variyam and Cawley \cite{variyam2006nutrition} and Loureiro $et$ $al.$ \cite{loureiro2012effects}, who dichotomized label use as $sometimes$ or higher and found significant health benefits of increased label use. We showed that a significant benefit of reading nutritional labels comes only with an $often$ or $always$ frequency, relative to reading at a $rare$ frequency. Such a conclusion could not be reached after dichotomizing the exposure or by other previously proposed methods. In fact, we estimated the treatment effect in our data set after dichotomizing label use into $sometimes$ or higher and $rare$ or $never$ levels. Under E1 elimination mechanism, and using subclassification on the propensity score with $K$ = 15 subclasses followed by regression adjustment, the estimated effect was not significant at the 0.05 nominal level (-0.05, 95\% CI, -0.29, 0.19). Although the direction of this effect was similar to our findings, this analysis did not capture the potential benefits of reading labels frequently. We recommend that policies and instructions for label use be updated to specify the extent with which one needs to read labels to reap the health benefits of a lower BMI. 

Subclassification on a GPS requires two assumptions, SUTVA and strong unconfoundedness. In our study, both assumptions seem reasonable given the design of the NHANES and the large number of observed covariates which were sufficiently balanced within each subclass, however the true validity of both of assumptions is unknown. Sensitivity approaches have been developed for binary treatment effects (see \cite{rosenbaum2004design, daniels2008missing, hosman2010sensitivity, liu2013inference}, to name a few), and a useful area for further research would examine the validity of these assumptions with an ordinal treatment. Further, because the NHANES is not a random sample, but a stratified random sample, our treatment effects generalize specifically to the population created by the sample; see \cite{hernan2008observational} and \cite{pearl2012external} for related discussions on the generalizability of observational data.

Inference using propensity scores is a preferred method of answering causal questions for comparative effectiveness research (CER), but generalizations of propensity scores to the multiple treatment setting are limited \cite{johnson2009good, rubin2010limitations}. The balance and estimation procedures provided here are important extensions of propensity score analysis to causal effects estimation for observational studies when the exposure is ordinal. These procedures yield, under proper assumptions, unbiased and transitive estimates of average treatment effects. 

\parskip=0.5\baselineskip \advance\parskip by 0pt plus 0.5pt

\section*{Acknowledgements} 

MJ Lopez was supported by the National Institute of Health (grant number R25GM083270). The authors would like to thank the anonymous reviewers for their comments and suggestions.

\section*{Conflict of Interest Statement} 

None declared

\bibliographystyle{vancouver}
\bibliography{ReferencesER}

\pagebreak

\begin{table*} 
\centering
\caption{Covariates \& Kendall's $\tau$ with nutritional label use}
\label{Kendall}
\begin{tabular}{l c r r}
\toprule
Variable & Type &\multicolumn{1}{c}{Kendall's $\tau$} & $p$-value\\
\midrule 
Gender, male & Binary & $-$0.19&\textless 0.001\\
Race, Hispanic& Binary &$-$0.14&\textless 0.001\\
Household size& Numeric&$-$0.13&\textless 0.001\\
Born to be fat?& Ordinal& $-$0.07&\textless 0.001\\
Drug user & Binary&$-$0.05&\textless 0.001\\
Smoker & Binary &$-$0.04&0.003\\
Safe sex & Binary&$-$0.01&0.338\\
Race, black&Binary & 0.00&0.867\\
Heart disease & Binary&0.00&0.816\\
Drinks per day& Numeric&0.00&0.699\\
Race, other& Binary& 0.01&0.292\\
Pregnant & Binary &0.01&0.430\\
$(\text{Meals away from home})^2$ & Numeric &0.02&0.060\\
Meals away from home & Numeric &0.02&0.048\\
Prior BMI & Numeric &0.04&0.001\\
Age & Numeric &0.06&\textless 0.001\\
Diabetic medicine& Binary&0.07&\textless 0.001\\
Diabetic & Binary&0.10&\textless 0.001\\
Race, white & Binary&0.11&\textless 0.001\\
Doct. advice 2 (reduce weight for chol.) &Binary &0.11&\textless 0.001\\
Doct. advice 3 (less fat for disease risk) & Binary&0.11&\textless 0.001\\
Income & Ordinal&0.12&\textless 0.001\\
Weight thoughts & Ordinal &0.12&\textless 0.001\\
Food security & Ordinal &0.13 & \textless 0.001\\
Doct. advice 1 (less fat for chol.) & Binary &0.13&\textless 0.001\\
Doct. advice 4 (reduce weight for disease risk) &Binary & 0.13&\textless 0.001\\
Healthy diet & Binary &0.16&\textless 0.001\\
$(\text{Metabolic Equivalence})^2$ & Numeric & 0.16 & \textless 0.001\\ 
Metabolic equivalence  & Numeric&0.18&\textless 0.001\\
Heard of diet guidelines& Binary&0.24&\textless 0.001\\
Heard of 5-a-day program & Binary&0.24&\textless 0.001\\
Education & Ordinal &0.25&\textless 0.001\\
Heard of food pyramid&Binary& 0.28&\textless 0.001\\
\bottomrule
\end{tabular}
\end{table*}
 
\clearpage

\begin{table*}
\centering
\caption{Label Use by \emph{Drug User}, subclass 2, $K$ = 10}
\label{druguser}
\begin{tabular}{l c c c c c}
\toprule
\emph{Drug user} &Never & Rare & Some & Often & Always\\
\midrule 
Yes &11 & 7&15 & 7&7 \\
No & 148 &48 &85 &54&36 \\
\bottomrule
\end{tabular}
\end{table*}

\clearpage

\begin{table*}
\centering
\caption{Proportion of significant ($p$-value $\textless$ $\alpha$) within-subclass balance tests}
\label{taustab}
\begin{tabular}{c c c c c}
\toprule
Elimination&$K$ (\# subclasses) & &$\alpha  = 0.01$ &$\alpha  = 0.05$ \\
\midrule 
\multirow{3}{*}{E1} 	& 5 	& & 0.018 & 0.103 \\
			& 10 & 	& 0 	& 0.052 \\
			& 15 	& & 0.004 & 0.042 \\ \midrule
\multirow{3}{*}{E2} 	& 5 	& & 0.012 & 0.115 \\
			& 10 	& & 0.009	& 0.079 \\
			& 15 	& & 0.006 & 0.053 \\ \midrule
\multirow{3}{*}{E3} 	& 5 	& & 0.018 & 0.097 \\
			& 10 	& & 0.006	& 0.064 \\
			& 15 	& & 0.014 & 0.048 \\
\bottomrule
\end{tabular}
\end{table*}

\tiny
\begin{landscape}
\centering
\begin{table*}
\scriptsize
\caption{$PATE$ estimates (standard errors in parenthesis) of BMI due to increased nutritional label use}
\label{ATE}
\begin{tabular}{c c c p{1.1cm} p{1.1cm} p{1.1cm} p{1.2cm} p{1.1cm} p{1.1cm} p{1.2cm} p{1.1cm} p{1.2cm} p{1.2cm}} 
\toprule
Elim. & $K$ & Adjust. &Rare v Never & Some v Never  & Often v Never&Always v Never & Some v Rare& Often v Rare &Always v Rare & Often v Some & Always v Some & Always v Often\\
\midrule 

\multirow{8}{*}{E1} &\multirow{4}{*}{10}
&A1 	&0.30 (0.18)&0.13 (0.14)&$-$0.09 (0.16)&$-$0.09 (0.17)&$-$0.17 (0.17)&$-$0.39 (0.19)**&$-$0.39 (0.20)&$-$0.22 (0.16)&$-$0.22 (0.16)&0.00 (0.18)\\
& &A2 	& 0.26 (0.18)&0.19 (0.15)&$-$0.07 (0.17)&$-$0.08 (0.17)&$-$0.07 (0.18)&$-$0.33 (0.20)&$-$0.35 (0.20)&$-$0.26 (0.16)&$-$0.27 (0.17)&$-$0.01 (0.19)\\
 &\multirow{4}{*}{15}
&A1 	& 0.29 (0.18)&0.17 (0.14)&$-$0.08 (0.17)&0.01 (0.17)&$-$0.12 (0.17)&$-$0.36 (0.19)&$-$0.28 (0.20)&$-$0.24 (0.16)&$-$0.16 (0.17)&0.08 (0.19)\\
& &A2 	& 0.25 (0.19)&0.17 (0.15)&$-$0.09 (0.17)&$-$0.03 (0.18)&$-$0.07 (0.18)&$-$0.34 (0.20)&$-$0.27 (0.21)&$-$0.26 (0.17)&$-$0.20 (0.17)&0.06 (0.19)\\

\midrule 
\multirow{8}{*}{E2}  &\multirow{4}{*}{10}
&A1 	&0.33 (0.18)&0.14 (0.14)&$-$0.10 (0.16)&$-$0.08 (0.17)&$-$0.20 (0.17)&$-$0.43 (0.19)**&$-$0.41 (0.19)**&$-$0.23 (0.15)&$-$0.21 (0.16)&0.02 (0.18)\\
& &A2 	& 0.30 (0.18)&0.17 (0.15)&$-$0.10 (0.17)&$-$0.10 (0.17)&$-$0.13 (0.17)&$-$0.40 (0.19)**&$-$0.39 (0.20)&$-$0.27 (0.16)&$-$0.26 (0.17)&0.00 (0.18)\\
 &\multirow{4}{*}{15}
&A1 	&0.32 (0.18)&0.12 (0.14)&$-$0.08 (0.17)&$-$0.05 (0.17)&$-$0.20 (0.17)&$-$0.41 (0.19)**&$-$0.37 (0.20)&$-$0.20 (0.16)&$-$0.17 (0.17)&0.04 (0.19)\\
& &A2 	& 0.28 (0.19)&0.14 (0.15)&$-$0.09 (0.17)&$-$0.08 (0.18)&$-$0.14 (0.18)&$-$0.37 (0.20)&$-$0.36 (0.21)&$-$0.22 (0.16)&$-$0.22 (0.17)&0.01 (0.19)\\

\midrule

\multirow{8}{*}{E3}  &\multirow{4}{*}{10}
&A1 	&0.39 (0.17)**&0.15 (0.14)&$-$0.08 (0.16)&0.00 (0.16)&$-$0.24 (0.16)&$-$0.47 (0.18)**&$-$0.40 (0.19)**&$-$0.23 (0.15)&$-$0.15 (0.16)&0.08 (0.17)\\
& &A2 	& 0.36 (0.18)**&0.19 (0.14)&$-$0.04 (0.16)&$-$0.02 (0.17)&$-$0.17 (0.17)&$-$0.40 (0.18)**&$-$0.38 (0.19)**&$-$0.23 (0.15)&$-$0.22 (0.16)&0.01 (0.17)\\

 &\multirow{4}{*}{15}
&A1 	&0.33 (0.18)&0.15 (0.14)&$-$0.08 (0.16)&0.01 (0.17)&$-$0.18 (0.17)&$-$0.40 (0.18)**&$-$0.32 (0.19)&$-$0.22 (0.15)&$-$0.14 (0.16)&0.09 (0.17)\\
& &A2 	& 0.25 (0.18)&0.21 (0.14)&$-$0.10 (0.16)&$-$0.03 (0.17)&$-$0.04 (0.17)&$-$0.35 (0.19)&$-$0.28 (0.20)&$-$0.31 (0.15)**&$-$0.24 (0.16)&0.06 (0.18)\\

\midrule 
\multicolumn{3}{l}{SBC (as in \cite{drichoutis2005nutrition})} &-0.04 (0.69) & 0.95 (0.43)** & 0.60 (0.54) & 0.13 (0.65) & -0.45 (0.55) & 0.79 (0.67) & 0.54 (0.69) & 0.34 (0.41) & -0.63 (0.51) & -0.07 (0.48)\\ 
\multicolumn{3}{l}{IPTW (as in \cite{feng2012generalized}, E1)} &0.49 (0.42) & 0.39 (0.31) & -0.10 (0.35) & 0.53 (0.42) & -0.08 (0.41) &-0.58 (0.42)
	& 0.05 (0.47) & -0.50 (0.29) & 0.14 (0.45) & 0.63 (0.43)\\
\multicolumn{3}{l}{IPTW (as in \cite{feng2012generalized}, E2)} &0.52 (0.46) & 0.29 (0.31) & -0.24 (0.36) & 0.53 (0.43) & -0.23 (0.46) &-0.76 (0.43)
	& 0.01 (0.49) & -0.53 (0.34) & 0.24 (0.44) & 0.77 (0.41)\\
\multicolumn{3}{l}{IPTW (as in \cite{feng2012generalized}, E3)} &0.59 (0.41) & 0.61 (0.31) & -0.01 (0.31) & 0.54 (0.38) & 0.02 (0.46) &-0.61 (0.36)
	& -0.06 (0.42) & -0.62 (0.29)** & -0.08 (0.40) & 0.55 (0.38)\\  \midrule
\multicolumn{12}{l}{E1: No subject elimination}\\ 
\multicolumn{12}{l}{E2: Eliminate subjects with extreme linear predictors}\\ 
\multicolumn{12}{l}{E3: Eliminate subjects with extreme continuous $X$ or extreme linear predictors}\\ \midrule
\multicolumn{12}{l}{A1: X = selected covariates (See Section~\ref{BA})}\\ 
\multicolumn{12}{l}{A2: X = all covariates}\\ \midrule
\multicolumn{12}{l}{**Significant at 0.05 level}\\ \bottomrule
\end{tabular}
\end{table*} 
\end{landscape}

\tiny
\begin{landscape}
\begin{table*}
\footnotesize
\centering
\caption{Simulated coverage, bias, and standard deviation of bias of seven $PATE$ estimators using two hypothetical full data sets}
\label{ATESim}
\begin{tabular}{c l c c c >{\hfill}p{1.85cm} >{\hfill}p{1.85cm} >{\hfill}p{1.85cm} >{\hfill}p{1.85cm} } \\ \toprule
& & \multicolumn{2}{c}{$Coverage$ \%} & &\multicolumn{4}{c}{ Mean $bias$ (SD) by label use comparison, vs. $Never$} \\ \cline{3-4} \cline{6-9}
PATE \rule{0pt}{12pt}  & Estimator    & $Average$*  & $Complete$**& & $Rare$ & $Some$ & $Often$ &$Always$ \\ \midrule

\multirow{7}{*}{Set 1 $PATE_{t,s} = 0$} 
& Subclass only, K=5		& 0.91 & 0.60 & & 		0.18 (0.41) & 0.12 (0.34) 		& $-$0.07 (0.35)	& 0.14 (0.34)  \\
& Subclass w/ Regression, K=5& 0.91 & 0.72  & & 		0.01 (0.17) & $-$0.01 (0.13) 	& 0.01 (0.13)		& $-$0.01 (0.15)   \\
& Subclass only, K=15		& 0.95 & 0.59 & & 		0.15 (0.43) & 0.08 (0.35) 		& $-$0.12 (0.35) 	& 0.08 (0.34) \\
& Subclass w/ Regression, K=15 & 0.96 & 0.74  & & 		0.00 (0.17) & $-$0.00 (0.14) 	& 0.00 (0.14) 		& 0.02 (0.16) \\
& Naive difference in means	 	& 0.77 & 0.22 & & 	0.45 (0.40) & 0.50 (0.38) 		& 0.44 (0.45) 		& 0.66 (0.48)\\
& Standard regression			& 0.94 & 0.70 & & 	0.01 (0.16) & 0.00 (0.14)		& 0.00 (0.14) 		& $-$0.00 (0.14)\\
& IPTW			& 0.88 & 0.57 & & 				0.67 (0.36) & 0.46 (0.41) 		& 0.06 (0.49) 		& 0.58 (0.62)  \\
\midrule 
\multirow{7}{*}{Set 2  $PATE_{t,s} \ne 0$}
& Subclass only, K=5		& 0.97& 0.80 	& &0.08 (0.39)		& 0.05 (0.28)	& 0.06 (0.31)	& 0.06 (0.35)  \\
& Subclass w/ Regression, K=5 & 0.96&  0.80 	& & 0.03 (0.38)		& $-$0.02 (0.27)& 0.03 (0.31)	& 0.03 (0.34) \\
& Subclass only, K=15		& 0.96& 0.80 	& &0.06 (0.41)		& 0.02 (0.30)	& 0.00 (0.32)	& 0.04 (0.37)\\
& Subclass w/ Regression, K=15  & 0.97& 0.78 & & 0.03 (0.43)		& $-$0.01 (0.30)& 0.02 (0.34)	& 0.04 (0.37)\\
& Naive difference in means	 & 0.83& 0.26 	& &0.29 (0.36)		& 0.41 (0.28)	& 0.73 (0.31)	& 0.24 (0.34)  \\
& Standard regression	& 0.90& 0.52 		& & $-$0.01 (0.36)	& 0.05 (0.25)	& 0.21 (0.27)	& $-$0.31 (0.31)\\
& IPTW			& 0.76 & 0.30 			& & 0.64 (0.37)		& 0.81 (0.33)	& 1.93 (0.36)	& 1.03 (0.40) \\ 
\midrule 
\multicolumn{9}{l}{* Fraction of all $\widehat{PATE}$ intervals containing the true $PATE$}\\ 
\multicolumn{9}{l}{** Fraction of simulations with all 10 pairwise $\widehat{PATE}$ intervals containing the true $PATE$}\\ \bottomrule
\end{tabular}
\end{table*}
\end{landscape}

\appendix

\normalsize
\section{Proof of unbiasedness}\label{proof}

Here we show that $\widehat{PATE}_{k(t,s)}$ is unbiased for $PATE_{k(t,s)}$. With $Y_{ik}(t)|\boldsymbol{X_{ik}} \sim N(\mu_{ikt},\sigma^2)$ as in Equation \ref{RM}, we model $Y_{ik}|\left\{\boldsymbol{X_{ik}},T\right\}$ within each subclass, where 
\begin{eqnarray}  \mu_{ikt} &=& \sum_{t=1}^Z \alpha_{kt}I_t(T_i=t) + \boldsymbol{{\gamma_k}X_{ik}} \nonumber \\
	&=& \alpha_{k1}I(T_i=1) + .. \alpha_{kZ}I(T_i=Z) + \boldsymbol{{\gamma_k}X_{ik}}  \nonumber \\
\text{and } \widehat{PATE}_{k(t,s)} &=& \hat{\alpha}_{kt} - \hat{\alpha}_{ks} \nonumber   \end{eqnarray}

\noindent Using $\hat{\alpha}_{kt}$ and $\hat{\alpha}_{ks}$, the maximum likelihood estimates of $\alpha_{kt}$ and $\alpha_{ks}$, we have $E[\hat{\alpha}_{kt} - \hat{\alpha}_{ks}] = E[\alpha_{kt} - \alpha_{ks}]$ \cite{myers1990classical}. 

Next we show $E[\alpha_{kt} - \alpha_{ks}] = PATE_{k(t,s)}$. As in Section 3.1, we assume the covariate distribution within each subclass is equal in expectation between those at different doses, that \begin{eqnarray}
E[\boldsymbol{X_{ik}}|T=t] = E[\boldsymbol{X_{ik}}|T=s] \qquad \qquad \qquad \qquad \qquad \text{(App. 1)} \nonumber \end{eqnarray} 
\noindent By properties of the Normal distribution, $E[Y_{ik}|\boldsymbol{X_{ik}},T=t] = \alpha_{kt} + \boldsymbol{{\gamma_k}X_{ik}}$ and $E[Y_{ik}|\boldsymbol{X_{ik}},T=s] = \alpha_{ks} + \boldsymbol{{\gamma_k}X_{ik}}$, thus: \begin{eqnarray}
E[\alpha_{kt} - \alpha_{ks}] &=& E[\alpha_{kt}] - E[\alpha_{ks}] \nonumber \\
	&=& E [ E[Y_{ik}(t)|\boldsymbol{X_{ik}},T=t] ] - E[ [Y_{ik}(s)|\boldsymbol{X_{ik}},T=s] ] \qquad  \text{(by App. 1)}\nonumber  \\
	&= &E[ E[Y_{ik}(t)|\boldsymbol{X_{ik}}]] - E[[Y_{ik}(s)|\boldsymbol{X_{ik}}] ]  \qquad \qquad \qquad \qquad \text{(by unconfoundedness)} \nonumber \\
	&= &E[ Y_{ik}(t)] - E[Y_{ik}(s)]  \nonumber \\
	&= & PATE_{k(t,s)}  \nonumber \\ \nonumber 
 \end{eqnarray}
\newpage

\section{Figures}\label{figsec}

\vspace{5 mm}
\centering ***************************Figure 4 here***************************\\ 
\vspace{5 mm}
\begin{em} \footnotesize
Histograms of $z$-test statistics of statistical dependency between covariates and label use using Kendall's $\tau$, for $K$ subclasses and subject elimination E1-E3 \end{em}
\vspace{5 mm}

\begin{description}
\item[E1:] No subject elimination, n = 4272
\item[E2:] Eliminate subjects with extreme linear predictors, n = 4142
\item[E3:] Eliminate subjects with extreme continuous $X$ or extreme linear predictors, n = 4076
\end{description}
\newpage

\section{Tables}

\subsection{Study population \& those excluded} 
\label{Compare}

\vspace{5 mm}

\centering
This table gives study characteristics of subjects included and excluded from our study for having missing covariate values (\% shown unless otherwise indicated)

\vspace{5 mm}

\begin{tabular}{l l c c}
\toprule
\multirow{2}{*}{Covariate} & \multirow{2}{*}{Description}  & In study & Eliminated  \\ 
  &    & n = 4272 & n = 372  \\ 
\midrule 
Age &mean (SE) &  47.3 (18.5) & 53.2 (20.8)\\
BMI &mean (SE) & 28.8 (6.8) & 28.7 (6.4)\\
Metabolic equivalence  &mean (SE) &8.6 (12.1) & 7.9 (4.5)\\
Diabetic&  &74 & 74\\
Drug user && 8& 5\\
Heard of diet guidelines&  & 43& 29\\ 
Gender & Males & 48 & 48\\\midrule
\multirow{5}{*}{Nutritional label use} & Never & 32 & 45\\
& Rare & 10& 10\\
& Some&   22& 20\\
& Most of the time& 19 & 12\\
& Always& 17 & 14\\ \midrule
\multirow{4}{*}{Race} & Hispanic & 22 & 39\\
 & white & 51 & 39\\
 &  black &   23& 19\\
 & other & 4 & 3\\
\bottomrule
\end{tabular}

\pagebreak

\subsection{Covariates not included in propensity score model}
\label{Extra}

\vspace{10 mm}

\centering
This table shows variables not included in our propensity score model (which were eventually balanced on through subclassification), and their original Kendall's $\tau_b$ correlation with label use 

\vspace{10 mm}

\begin{tabular}{l c r c}
\toprule
Variable & Description &\multicolumn{1}{c}{Kendall's $\tau_b$} & $p$-value\\
\midrule 
Blood pressure problems & Binary&0.07 \text{    }&\textless 0.001\\
Cocaine use& Binary & $-$0.01 \text{    }&0.60\\
Marijuana use & Binary & $-$0.05 \text{    } & \textless 0.001\\
Marital status (Yes vs. No)& Binary  &0.03 \text{    }&0.03\\
Ever drink 5+ drinks per day& Numeric&$-$0.06 \text{    }&\textless 0.001\\
Weight control&Binary & 0.11 \text{    }&\textless 0.001 \\
\bottomrule
\end{tabular}

\begin{landscape}
\subsection{Covariates used \& a brief description}
\label{NHANES}\centering
\scriptsize
\begin{tabular}[p]{p{1in} p{1.5in} p{4.5in}}
\toprule
Type & Variable Name & Description/Levels \\
\midrule 
\multirow{6}{*}{\parbox{2cm}{Numeric}}  & Age & Years of respondent\\
 &Drinks per day& \# of alcoholic drinks consumed per day over the past 12 months\\
 & Household size& \# of people in household\\
 &Meals away from home& \# weekly meals prepared outside of home\\
 &Metabolic equivalence & Total metabolic activity rate \\
 &Prior BMI & Calculated using respondent's estimate of their weight from one-year ago and their current height\\ \midrule
\multirow{6}{*}{\parbox{2cm}{Ordinal}} & Born to be fat?&Are people born to be fat? Respondent answers: strongly disagree, 
	somewhat disagree, neither agree nor disagree, somewhat agree, or strongly agree\\
&Education & HS/GED, some college or associate's degree, or college graduate\\
& Food security & Household food security: low, marginal, or full\\
& Income & Household income: Less than \$24,999/yr, between \$25,000 and \$54,999/yr, or greater than \$55,000/yr\\
& Weight thoughts & Respondent's thoughts on his or her own weight: underweight, about the right weight, or overweight \\
\midrule
\multirow{17}{*}{\parbox{2cm}{Nominal} }&Race& Hispanic, non-Hispanic white (white), non-Hispanic black (black), Other\\
& Diabetic & Respondent has been told by a doctor of diabetes or pre-diabetic conditions\\
& Diabetic medicine & Respondent takes insulin or pills for diabetes\\
& Doct. advice 1& Doctor's advice to respondent: eat less fat for cholesterol\\
& Doct. advice 2& Doctor's advice to respondent: reduce weight for cholesterol\\
& Doct. advice 3& Doctor's advice to respondent: eat less fat for disease risk \\
& Doct. advice 4& Doctor's advice to respondent: reduce weight for disease risk\\
& Drug user & Respondent has used hashish, marijuana, cocaine, heroin, or methamphetamine in the past month \\
& Gender &Male, female\\
& Healthy diet & Respondent rates diet as good or better\\
& Heard of 5-a-day program & Respondent has heard of 5-a-day program\\
& Heard of diet guidelines& Respondent has heard of diet guidelines\\
& Heard of food guide pyramid& Respondent has heard of food guide pyramid\\
& Heart disease & Respondent suffers from coronary heart disease, stroke, or liver condition\\
& Pregnant &Respondent is pregnant\\
& Safe sex & Respondent has not had sex without a condom in the past year\\
& Smoker & Respondent smokes cigarettes\\
\bottomrule
\end{tabular}
\end{landscape}

\end{document}